\documentclass[fleqn,twoside]{article}
\usepackage{espcrc2}
\usepackage{graphicx}
\usepackage[figuresright]{rotating}
\def\csw{c_{\rm sw}}

\title{\hfill{ \normalsize JLAB-THY-01-38}\\[1.0ex]
Excited nucleon spectrum using a non-perturbatively improved
  clover fermion action}

\author{D.G.~Richards\address{Jefferson Laboratory, MS 12H2, 12000
Jefferson Avenue, Newport News, VA 23606, USA}
\address{Department of Physics, Old Dominion University, Norfolk, 
        VA 23529, USA} (LHPC Collaboration)\\[1.0ex]
M.~G\"{o}ckeler\address[REGEN]{Institut f\"ur Theoretische Physik, 
Universit\"at Regensburg, D-93040 Regensburg, Germany}, 
R.~Horsley\address[DESY]{Deutsches Elektronen-Synchrotron DESY,
             John von Neumann Institute for Computing NIC/Deutsches
	     Elektronen-Synchrotron DESY,
                    D-15738 Zeuthen, Germany}
\address{Institut f\"ur Physik, Humboldt-Universit\"at zu Berlin,
                    D-10115 Berlin, Germany},
D.~Pleiter\addressmark[DESY], 
P.E.L.~Rakow\addressmark[REGEN], 
G.~Schierholz\address{Deutsches Elektronen-Synchrotron DESY,
                    D-22603 Hamburg, Germany}\addressmark[DESY] (QCDSF
Collaboration)\\[1.0ex]
C.M.~Maynard\address{Department of Physics \& Astronomy, University of
  Edinburgh, Edinburgh EH9~3JZ, Scotland, UK} (UKQCD Collaboration)}
\begin{document}
\begin{abstract}
We discuss the extraction of negative-parity baryon masses from
lattice QCD calculations.  The mass of the lowest-lying
negative-parity $J = 1/2^{-}$ state is computed in quenched lattice
QCD using an ${\cal O}(a)$-improved clover fermion action, and a
splitting found with the nucleon mass. The calculation is performed on
two lattice volumes, and three lattice spacings enabling a study of
both finite-volume and finite-lattice-spacing uncertainties. A
measurement of the first excited radial excitation of the nucleon
finds a mass considerably larger than that of the negative-parity
ground state, in accord with other lattice determinations but in
disagreement with experiment.  Results are also presented for the
lightest negative-parity $I=3/2$ state.
\end{abstract}
\maketitle

\section{INTRODUCTION}
The $N^*$ spectrum exhibits many features that are emblematic of QCD;
that they contain three valence quarks is a manifestation of SU(3)
symmetry, while the linearly rising Regge trajectories are evidence
for a flux-tube-like confining force.  Finally, the fine and hyperfine
structure reveals details of the interquark interactions.  Thus the
study of the spectrum of excited nucleon resonances has always been a
vital component of the experimental hadronic physics program, as
typified by the Hall B experimental effort at Jefferson Laboratory.

The motivation for studying the $N^*$ spectrum in many ways mirrors
that for the hybrids, which also provide an arena in which to explore
flux tubes.  One distinction is that, in the case of nucleons, all
spin-parity combinations are accessible to the quark model, and thus
there are no nucleon ``spin exotics''.  However, both hybrids and
excited nucleons share the feature that they are sensitive to the
presence of excited glue.

The observed excited nucleon spectrum already poses many questions.
There are ``missing resonances'', predicted by the quark model but not
yet observed experimentally.  Most strikingly, there is the so-called
Roper resonance, and the lightest negative-parity $\Lambda$ state, the
$\Lambda(1405)$ with anomalously light masses suggesting that they may
be ``molecular'' states, rather than true three-quark resonances.
Lattice gauge theory calculations have a vital role to play in
resolving these questions.

In an attempt to address these issues, there have been several recent
calculations of the lowest-lying excited nucleon masses, emphasising
in particular the extent to which the parity partners are accessible
to lattice calculation.  The first calculation employed the
highly-improved $D_{234}$ fermion action\cite{lee98,lee00}, whilst a
second calculation employed domain-wall
fermions\cite{sasaki99,sasaki01}.  Both calculations exhibited a clear
splitting between the masses of the $N^{1/2+} (937)$ and
$N^{1/2-}(1535)$ states, with an indication that the lowest-lying
radial excitation of the nucleon has a mass considerably larger than
that of the $N^*(1535)$, casting doubt on the nature of the Roper
resonance.

In this talk, we present a calculation of the lowest lying negative
parity nucleon mass using an ${\cal O}(a)$-improved
Sheikholeslami-Wohlert (SW), or clover, fermion action; preliminary
results were presented in ref.~\cite{lat00}, and a comprehensive
analysis appeared in ref.~\cite{lhpc01}.  By choosing the coefficient
of the improvement term appropriately, all ${\cal O}(a)$
discretisation uncertainties are removed, ensuring that the continuum
limit is approached with a rate proportional to $a^2$.  For a subset
of our lattices, masses of the first radial excitation of the nucleon
and of the parity partner of the $\Delta$ are presented.

The calculation of the excited nucleon spectrum places particularly
heavy demands on lattice spectroscopy.  The excited nucleon states are
expected to be large; the size of a state is expected to double with
each increase in orbital angular momentum.  Thus a lattice study of
the excited nucleon spectrum requires large lattice volumes, with
correspondingly large computational requirements.  Furthermore, the
states are relatively massive, requiring a fine lattice spacing, at
least in the temporal direction.  These requirements could be
satisfied with much greater economy using the clover fermion action
than using the domain-wall or overlap formulation.  Thus it is
important to establish that the negative parity states are indeed
accessible to calculations using the clover action.  Finally, by
comparing the masses obtained using the clover action with a
calculation, at a single quark mass, using the Wilson fermion action,
we also gain insight into the nature of the interaction responsible for the
splitting in the parity doublet.

The layout of the remainder of this talk is as follows.  In the next
section, I shall describe the construction of hadronic operators, and
in particular the projection onto positive- and negative-parity states,
and summarises other computational details. Section~\ref{sec:results}
contains the results for the masses of the lowest-lying positive- and
negative-parity nucleon using the SW fermion action, and for a subset
of the lattices the masses obtained for the Roper resonance.  The talk
concludes with a discussion, a comparison with other lattice and
phenomenological calculations, and prospects for future studies.

\section{SIMULATION DETAILS}\label{sec:operators}
\subsection{Baryon operators}
Historically, there has been relatively little study of the
negative-parity baryon states, though there had been numerous
computations of the nucleon radial excitations.  It is therefore
useful to review the r\^{o}le that parity plays in the construction of
baryon operators, and how this r\^{o}le is manifest in the case of
lattice calculations.  We illustrate the discussion through
consideration of the usual nucleon interpolation operators:
\begin{eqnarray}
N_1^{1/2+} & = & \epsilon_{ijk} (u_i^T C \gamma_5 d_j) u_k\label{eq:N1},\\
N_2^{1/2+} & = & \epsilon_{ijk} (u_i^T C d_j) \gamma_5 u_k\label{eq:N2},\\
N_3^{1/2+} & = & \epsilon_{ijk} (u_i^T C \gamma_4 \gamma_5 d_j) u_k.
\label{eq:N3}
\end{eqnarray}
These operators have an overlap with particle states of both positive and
negative parities.  In order to construct operators of definite
parity, we introduce the parity-projection operator:
\begin{equation}
P = (1 \pm \gamma_4),
\end{equation}
and construct the correlators
\begin{equation}
C_{\pm}(t) = \sum_{\vec{x}} \langle 0 | T N(\vec{x}, t) (1 \pm \gamma_4)
\overline{N}(0) | 0 >,
\end{equation}
where $N$ is a baryon interpolating operator.  On a periodic or
anti-periodic lattice, the two time orderings result in forward
propagating states of positive (negative) parity, and backward
propagating states which are anti-particles of negative (positive)
parity states if the projection operator is chosen with positive
(negative) sign.  Thus the best delineation that may be constructed is
between a forward propagating particle, and the backward propagating
anti-particle of the parity partner.  At large distances, where $t \gg
1$ and $N_t - t \gg 1$, the correlators assume the time behaviour
\begin{eqnarray}
C_{+}(t) & \rightarrow & A^+ e^{-m^+ t} + A^- e^{-m^-(N_t
- t)} \label{eq:corr_pos}\\
C_{-}(t) & \rightarrow & A^- e^{-m^- t} + A^+ e^{-m^+(N_t -
t)}\label{eq:corr_neg}
\end{eqnarray}
where $m_i^+$ and $m_i^-$ are the lightest positive- and
negative-parity masses respectively.

For the case of the nucleon ground state, the ``diquark'' part of both
$N_1$ and $N_3$ couples upper (large) spinor components, while that in
$N_2$ involves both an upper and lower spinor component and thus
vanishes in the non-relativistic limit\cite{lee98}.  Thus our
expectation is that the operators $N_1$, $N_3$ should give a better
overlap with the positive-parity ground state nucleon compared with
operator $N_2$, and this is confirmed in lattice simulations.  For
some of the lattices used in the calculation, the positive-parity
nucleon mass is obtained using the non-relativistic quark
operators~\cite{qcdsf95}, defined by
\begin{equation}
\psi \rightarrow \psi^{\rm NR} = \frac{1}{2} ( 1 + \gamma_4) \psi,\,
\overline{\psi}^{\rm NR} = \overline{\psi} \frac{1}{2} (1 + \gamma_4).
\end{equation}
corresponding to the ``large'' components in the non-relativistic
limit;  this results in a factor of two reduction in computational
effort at the expense of some loss of statistics.

\subsection{Simulation Details}
The calculation is performed in the quenched approximation to QCD.
The standard Wilson gluon action is employed, whilst the quark
propagators are computed using the Sheikholeslami-Wohlert clover
fermion action, using the non-perturbative coefficient of the clover
term determined in refs.~\cite{AlphaIII} and \cite{SCRI_imp}.  Thus
the hadron masses are free of all ${\cal O}(a^2)$ discretisation
errors.  

The propagators are computed using both local sources, and spatially
extended ``smeared'' sources, using either the fuzzed-source method
(F) or Jacobi-smeared sources (J) as described in
references~\cite{fuzzing} and~\cite{SFsmear} respectively.  The
simulation parameters are summarized in Tables~\ref{tab:params}.  The
errors on the fitted masses are computed using a bootstrap procedure.
Different numbers of configurations are used at different quark masses
at some of our $\beta$ values, precluding the use of correlated fits
in some of the chiral extrapolations.  Thus a simple uncorrelated
$\chi^2$ fit is performed, with the uncertainties computed from the
variation in the $\chi^2$.  Calculations of the light hadron spectrum
using these lattices were presented in references~\cite{UKQCD} and
\cite{QCDSF,pleiter00}.
\begin{table*}[htb]
\caption{The parameters of the lattices used in the calculation.  The
labels $J$ and $F$ refer to use of Jacobi and ``fuzzed'' quark sources
respectively.}\label{tab:params}
\begin{tabular}{cccccc}
\hline $\beta$ & $\csw$ & $L^3\cdot T$ & $L\,[{\rm fm}]$ & $\kappa$ &
Smearing \\ \hline 6.4 & 1.57 & $32^3\cdot 48$ & 1.6 & $0.1313, \,
0.1323,\, 0,1330, \, 0.1338, \, 0.1346, \, 0.1350$ & J\\[1.0ex] 6.2 &
1.61 & $24^3\cdot48$ & 1.6 & $0.1346,\,0.1351,\,0.1353$ & F \\ & &
$24^3\cdot 48$ & 1.6 & $0.1333,\, 0.1339, \, 0.1344,\, 0.1349, \,
0.1352$ & J \\ & & $32^3\cdot 64$ & 2.1 & $0.1352, \, 0.1353, \,
0.13555$ & J \\[1.0ex] 6.0 & 1.76 & $16^3\cdot48$ & 1.5 &
$0.13344,\,0.13417,\,0.13455$ & F\\ & & $16^3\cdot 32$ & 1.5 &
$0.1324,\, 0.1333, \, 0.1338, \, 0.1342$ & J\\ & & $24^3\cdot 32$ &
2.2 & $0.1342, \, 0.1346, \, 0.1348$ & J\\[1.0ex] \hline
\end{tabular}\\[1.0ex]
Lattice sizes in physical units are quoted using $r_0$
to set the scale~\protect\cite{wittig98}.
\end{table*}

The clover term removes the leading chiral-symmetry-breaking
effects from hadron masses at finite $a$.  Since the lack of
degeneracy between the positive- and negative-parity baryon states is
a consequence of the spontaneous breaking of chiral symmetry, we also
compare the measurements of the mass splitting at a single quark mass
with those obtained using the Wilson fermion action.

\section{RESULTS}\label{sec:results}
\begin{figure}
\includegraphics[width=205pt]{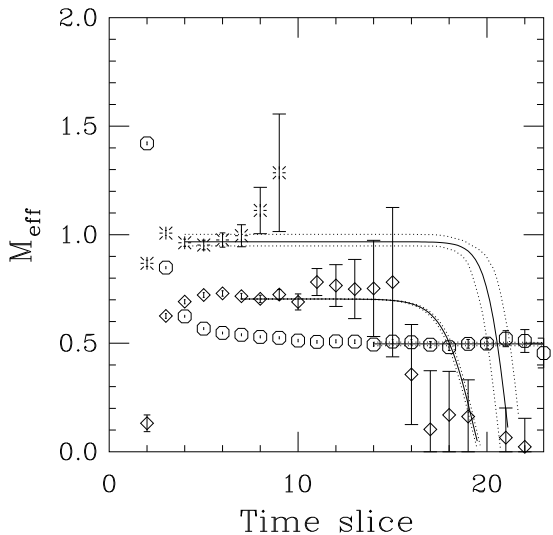}
\caption{The effective masses for the $N^{1/2+}$ channel (circles) and
  the $N^{1/2-}$ channel (diamonds) at $\beta = 6.2$, with $\kappa =
  0.1351$, using the interpolating operator $N_1$ of
  eqn~\protect\ref{eq:N1}. Also shown as the bursts is the effective
  mass of the positive-parity state determined using the
  interpolating operator $N_2$.}
\label{fig:eff_mass}
\end{figure}

The masses of the lowest-lying $N^{1/2+}$ and $N^{1/2-}$ states are
obtained from a simultaneous, four-parameter fit to the positive-
and negative-parity correlators $C_{+}(t)$ and $C_{-}(t)$
constructed from the ``good'' nucleon operator $N_1$ of
eqn.~\ref{eq:N1}.  A clear signal for the mass of the lightest
negative-parity state notably higher than that for the positive-parity
state is observed, and is illustrated for one of our ensembles in
Figure~\ref{fig:eff_mass}.

For the chiral extrapolation of the hadron masses, we adopt the ansatz
\begin{eqnarray}
(a m_X)^2 = (a M_X)^2 + b_2 (a m_{\pi})^2\label{eq:chiral_extrap},
\end{eqnarray}
where we use upper-case letters to denote masses obtained in the
chiral limit and $X$ is either $N^{1/2+}$ or $N^{1/2-}$.  We include
data at different volumes, but at the same $\beta$, in the chiral
extrapolations, but treat the fuzzed and Jacobi-smeared data
independently; they give slightly different masses, reflecting the
different contributions of the excited states in the two cases.  Note
that the fit to $(a m_X)^2$ is formally the same as one to $a m_X$ at
small pion mass, but in practice provides a notably better description
of the data.

\begin{figure}
\includegraphics[width=205pt]{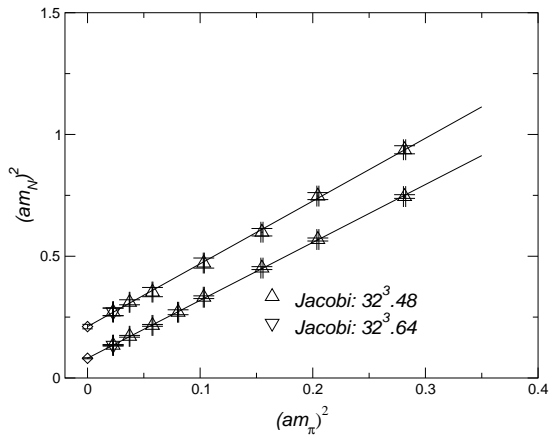}
\caption{The masses in lattice units of the lowest-lying positive- and
  negative-parity nucleons at $\beta = 6.4$.  The curves are from fits
  to $m_X^2$ using eqn.~\protect\ref{eq:chiral_extrap}.}
\label{fig:chiral_extrap}
\end{figure}

It is known that there are non-analytic terms contributing to the
chiral extrapolation of physical quantities, and the contribution of
the non-analytic terms arising from the pion-induced baryon self
energies has been explored in reference~\cite{leinweber99}.  In the
quenched approximation, the leading, non-analytic terms can assume a
different form.  From quenched chiral perturbation theory, one expects
leading non-analytic terms which are linear in
$m_{\pi}$~\cite{labrenz96}; the coefficient of the leading term is
predicted to be negative.  We therefore also attempted to fit the data
at $\beta = 6.4$ to the form
\begin{equation}
(am_X)^2 = (a M_X)^2 + b_1 (a m_{\pi}) + b_2 (a m_{\pi})^2.
\label{eq:non_analytic}
\end{equation}
Fitting the $N^{1/2+}$ state we found a positive value of $b_1$,
though with a very large error that would still accommodate a negative
value.  Thus it is unclear whether eqn.~\ref{eq:non_analytic} provides
a reliable form with which to extrapolate data obtained with quark
masses around that of the strange to the chiral limit, emphasising
once again the need for data at the smaller quark masses at which the
pion cloud emerges.  The chirally extrapolated masses obtained using
this procedure, with $b_1$ unconstrained, differ from those obtained
using eqn.~\ref{eq:chiral_extrap} by around 5\%.  In view of the
difficulties discussed above, and because on some of our ensembles we
have insufficient data points to perform a three-parameter fit, we
quote our final results from fits to eqn.~\ref{eq:chiral_extrap}.  An
example of such a chiral extrapolation for the data at $\beta = 6.4$
is shown in Figure~\ref{fig:chiral_extrap}.

Ideally, the forms eqn.~\ref{eq:chiral_extrap}
and~\ref{eq:non_analytic} should be applied to the baryons masses
obtained in the infinite volume limit.  Indeed, the talk by Stuart
Wright at this workshop emphasised the importance of correctly
accounting for finite-volume effects in chiral
extrapolations~\cite{wright01}.  Our data do not admit this procedure,
but an investigation of the masses of the negative-parity state
obtained on the smaller and larger of our lattices at $\beta = 6.0$
and $6.2$ suggests that the masses of the negative-parity state on the
larger lattices are smaller than those on the smaller lattices by
around 5\%.

\begin{figure}
  \includegraphics[width=205pt]{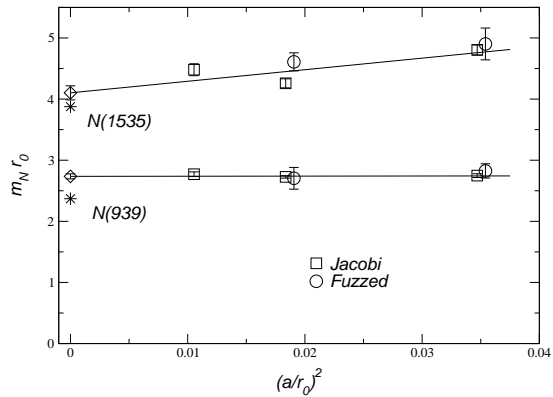}
  \caption{The masses of the lowest-lying positive- and negative-parity
    baryons in units of $r_0^{-1}$~\protect\cite{sommer94,wittig98}
    against $a^2$ in units of $r_0^2$.  The lines are linear fits in
    $a^2/r_0^2$ to the positive- and negative-parity baryon masses.  Also
    shown are the physical values.}
  \label{fig:continuum_extrap}
\end{figure}

In order to look at the discretisation uncertainties in our
data, we show in Figure~\ref{fig:continuum_extrap} the masses in units
of $r_0$ against the $a^2/r_0^2$, where $r_0 = 0.5~\mbox{fm}$ is the
hadronic scale~\cite{sommer94,wittig98}.  

\begin{figure}
\includegraphics[width=205pt]{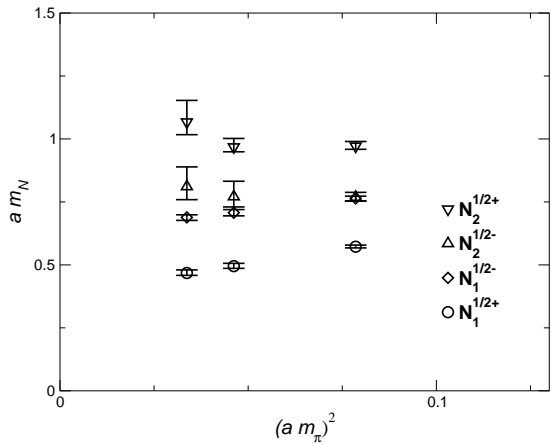}
\caption{The masses in lattice units of the lowest-lying $N^{1/2+}$
(circles) and the $N^{1/2-}$ (diamonds) obtained using $N_1$, and of
the lowest-lying $N^{1/2+}$ (down triangle) and $N^{1/2-}$ (up
triangle) using `$N_2$, obtained on the
$24^3\times 48$ lattices at $\beta = 6.2$.}
\label{fig:good_bad_chiral}
\end{figure}

On a subset of our lattices, we have extracted the ground-state
positive- and negative-parity masses using the operator $N_2$
of eqn.~\ref{eq:N2}.  The effective mass in the positive-parity
channel using this operator is shown in Figure~\ref{fig:eff_mass}; it
indicates an effective ground state mass considerably higher than that
of the lightest negative-parity state.  The signal for both the
positive- and negative-parity ground states degrades appreciably with
decreasing pion mass, as can be seen in
Figure~\ref{fig:good_bad_chiral} where we show the extracted
lowest-lying masses of both parities using each of the $N_1$ and $N_2$
interpolating operators.  A better determination of the first radial
excitation would require the measurement of the cross-correlator
between $N_1$ and $N_2$, as done in reference~\cite{sasaki01}.
However, the conclusion that the operator $N_2$ has a negligible
overlap with the ground state nucleon at these values of the
pseudoscalar mass seems inescapable.

\begin{figure}
\includegraphics[width=205pt]{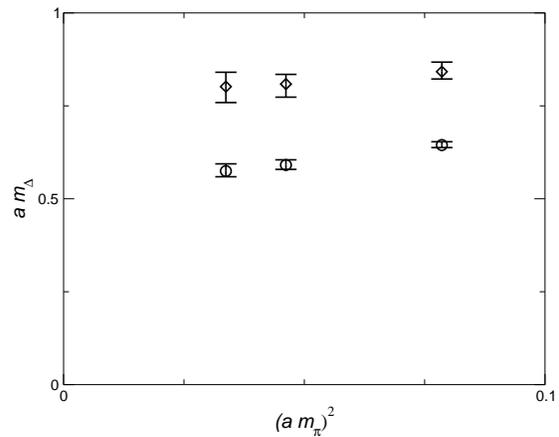}
\caption{The masses in lattice units of the $\Delta$ (circles) and its
  parity partner (diamonds) against the pseudoscalar mass obtained on
  the $24^3\times 48$ lattices at $\beta = 6.2$}
\label{fig:delta}
\end{figure}

The measurement of heavier excitations is inevitably noisier because of
the reduced signal-to-noise ratio.  On some of our lattices we have
measured the correlators for the $I = 3/2$ $\Delta$, using an
interpolating operator
\begin{equation}
\Delta^{3/2,1/2} = \epsilon_{ijk}(u^T_i C \gamma_{\mu} u_j) u_k.
\end{equation}
This has an overlap onto both spin-$3/2$ and spin-$1/2$ states, but
these can be distinguished using a suitable spin projection, and the
two parities delineated as described above.  The masses of the ground
state of both positive and negative parity are shown in
Figure~\ref{fig:delta}. Further success at determining the higher spin
states has been made in a recent determination
of the $N^{3/2}$ masses~\cite{lee01}.

\section{CONCLUSIONS}
We have seen that the use of the clover-improved fermion action
enables the delineation of the positive- and negative-parity excited
state masses, in accord with calculations using highly improved
Wilson, and domain-wall fermion actions.  On a subset of our lattices,
we also computed propagators using the unimproved Wilson fermion
action at a single quark mass, and found a parity mass splitting
consistent with that of the clover fermion action.  On reflection,
this is hardly surprising; under $SU(6)$ spin-flavour symmetry, the
low-lying positive-parity baryons can be assigned to an $l = 0$
multiplet, whilst the low-lying negative-parity baryons can be
assigned to an $l = 1$ multiplet.  Thus the mass-splitting can be
likened to the $P-S$ splitting in the meson sector, which is
well-known to be faithfully reproduced using the Wilson fermion
action.

The striking feature common to all the lattice calculations is the
apparent inversion of the ordering of mass of the lowest radial
excitation of the nucleon, and the mass of the parity partner of the
nucleon.  All the calculations, using different gluonic and fermionic
actions, suggest a splitting between the masses of the first
positive-parity radial excitation and the lightest negative-parity
state comparable to that between the masses of the negative-parity
state and the nucleon.  In contrast, the observed lightest radial
excitation of the nucleon is the Roper resonance with a mass of $1440
\rm{MeV}$, whilst the lightest negative-parity state is the heavier
$N^*(1535)$. We can see that lattice calculations of the spectrum are
already posing interesting questions about the nature of the Roper
resonance.

I have not discussed the issue of the nature of the $\Lambda(1405)$ in
this talk.  In the case of the positive-parity states, lattice
calculations, such as that of ref.~\cite{UKQCD}, suggest that the
quark mass behaviour of the baryon masses, in the range over which
they are measured, is well described by terms corresponding to the
``centres of gravity'' of the quark masses within the hadron.  This
expectation has been confirmed in ref.~\cite{adelaide02}, where the
lightest negative-parity $\Lambda$ resonance has a mass considerably
in excess of $1405~\mbox{MeV}$.  It should be noted, however, that a
recent large $N_C$ calculation has been able to accommodate a light
$\Lambda(1405)$~\cite{goity01}.

We have seen that lattice calculations are already addressing some of
the salient issues in baryon spectroscopy.  The calculation of the
radial excitations and the higher spin states will present further
challenges; the need to overcome the reduced signal-to-noise ratio
afflicting heavier excitations, the construction of suitable lattice operators
to isolate higher excitations, and the exploitation of improved data
analysis methods.  As these issues are tackled, we will be able to
tackle the more ambitious projects of transition form factors and
hadronic decays.

\section*{ACKNOWLEDGEMENTS}
This work was supported in part by DOE contract DE-AC05-84ER40150
under which the Southeastern Universities Research Association (SURA)
operates the Thomas Jefferson National Accelerator Facility, by EPSRC
grant GR/K41663, and PPARC grants GR/L29927 and GR/L56336.  CMM
acknowledges PPARC grant PPA/P/S/1998/00255. MG acknowledges financial
support from the DFG (Schwerpunkt ``Elektromagnetische
Sonden''). Propagators were computed using the T3D at Edinburgh, the
T3E at ZIB (Berlin) and NIC(J\"{u}lich), the APE100 at NIC (Zeuthen),
and the \textit{Calico} cluster at Jefferson Laboratory.

We are grateful for fruitful discussion with S.~Dytman, R.~Edwards,
J.~Goity, N.~Isgur, R.~Lebed, F.~Lee, C.~Michael, C.~Morningstar,
C.~Schat, A.~Thomas and S.~Wallace.

\end{document}